\documentclass[manuscript]{aastex}

\usepackage{epsfig}
\usepackage{subfigure}

\shorttitle{Modelling Pulsar Precession }

\shortauthors{Clifton \&  Weisberg}

\begin{document}

\title{A Simple Model for Pulse Profiles from Precessing 
Pulsars,    with Special Application to Relativistic Binary PSR B1913+16}

%\title{Pulse-Width Profiles from Precessing Binary Pulsars,
%    and the Likely Geometry of PSR1913+16}

\author{Timothy Clifton}
\affil{Department of Astrophysics, University of Oxford, Oxford OX1
  3RH, UK}
\email{T.Clifton@cantab.net}
\author{Joel M. Weisberg}
\affil{Department of Physics and Astronomy, Carleton College,
Northfield, MN 55057}
\email{ jweisber@carleton.edu}

\def\be{\begin{equation}}
\def\ee{\end{equation}}
\def\bea{\begin{eqnarray}}
\def\eea{\end{eqnarray}}
\def\l{\label}
\def\r{\ref}

%\usepackage{amssymb}
%\usepackage{amsmath}
%\usepackage{epsfig}
%\usepackage{subfigure}

%\date{{\normalsize {\today}}}
\received{2007 October 14}
\accepted{2008 February 4}

\slugcomment{Astrophysical Journal \textbf{679}, 687 (2008).}

\begin{abstract}

We study the observable pulse profiles that can be generated from
precessing pulsars.   A novel coordinate system is defined to aid
visualization of the observing geometry. Using this system we
explore the different families of profiles that can be generated by
 simple, circularly symmetric beam shapes.  An attempt is then made to
fit our model to the observations of relativistic binary PSR B1913+16.  It is found that while
qualitatively similar pulse profiles can be produced, this minimal
model is insufficient for an accurate match to the observational
data.  Consequently, we confirm that the emission beam of PSR B1913+16 must 
deviate from circular symmetry, as first reported by \citet{Weis}.
However, the approximate fits obtained suggest that it may be
sufficient to consider only minimal deviations from a 
circular beam in order to explain the data.  We also comment on the applicability
of our analysis technique to other precessing pulsars, both binary and
isolated.

\end{abstract}

\newpage

\section{Introduction}

Most pulsars  spin about an axis  that remains fixed in space,
relative to our line of sight.  However, in some
cases  the spin axis will precess.  The free precession of an isolated pulsar due to a 
body asymmetry with respect to its spin axis is one such case;  the general relativistic 
``geodetic'' precession of the spin axis of a pulsar in a binary system about its orbital 
angular momentum vector is another \citep{Dam, Bar1, Bar2, Borner, hari75}. 

Spin precession, whatever its cause,  allows our line of sight to progress
across the pulsar emission beam.  As this occurs,
the observed  pulse profile will also evolve.  The detection of
such changes in pulse shape is consequently a hallmark of spin axis precession. Whilst we are forever 
bound to receive beamed emission from a single latitude on  a non-precessing 
pulsar,  the presence of precession, and the resulting pulse shape changes, enable 
us to make inferences about 
the two-dimensional  structure of the pulsar emission beam, and the likely 
geometry of the system.

Secular pulse shape changes in PSR B1913+16 ascribed to geodetic precession
were first reported by \citet{wtr89}. As more data accumulated in the mid-1990s,
and the signature of precession became clearer,  it became possible to
use the observed change in the  separation of the two principal pulse
components \citep{Kramer} to pick the geometrical models that best fit
the observational data.  The pioneering work of \citet{Kramer} assumed a 
simple circular region emitting a
conical beam of radiation from the pulsar.  Using this
model Kramer performed a least-squares fit on the
observed pulse component separation data to find the best-fitting parameters 
that described the geometry of the system.  Later
investigations were performed by  Weisberg \& Taylor (2002, 2005; 
hereafter WT02, WT05).  In these papers, the authors decomposed
the pulse profile into symmetric and anti-symmetric parts and
considered not only the time evolution of the peaks 
of the symmetric pulse profile, but also the separations of the contours of lower intensity.  The
profiles they obtained show that the equal intensity contours in the
center of the profile appear to be moving together more quickly than
those at the edge, which may even be slowly moving
apart.  The authors were unable to account for this strange behavior with a
circular beam model; instead they
generalized the shape of the emission beam to that of an `hourglass' by 
adding two more parameters to the beam model.  
The parameters describing this
deformed shape, and those describing the geometry of the system, were
then fitted to the data to find a best model.

Our purpose here is to further investigate the observational consequences of a
precessing {\em{circular}} beam model. It is possible to
envisage a number of different ways in  which our
line of sight could precess through the pulsar's emission beam, each
leading to qualitatively different observations of the evolution of
the pulse profile here on Earth.  We investigate the
different `families' of profiles that can be generated by different
geometries of the pulsar.  Using this analysis we take a new look at
families that provide good candidates for the geometry of the
geodetically precessing binary pulsar PSR B1913+16, and use them to revisit the question 
of whether this simple model is adequate to fit the observed data, as
contested by WT02 and WT05.  We also comment on the applicability of this analysis
for other precessing pulsars.

\section{The Model}

In order to investigate the types of pulse profile variation that can be
observed from a precessing pulsar, we need to make
assumptions about the shape of the beam being emitted and
the nature of the precession process.
We will allow the emitting region to be two--dimensional - that is, we will 
consider an emitting {\em{area}} being projected outwards from 
the pulsar, rather than simply an emitting line.  This will allow us to 
generate families of pulse profiles of the flux density observed on 
Earth as a function of pulse longitude and precession phase.  For simplicity 
and clarity we will only consider
emission beams that are circularly symmetric.  More general beam shapes
could be considered, but this would unnecessarily
complicate the present study which is designed to focus on the effects of
different precession geometries and beam  orientations on the pulse profile.  
Furthermore,
for the purpose of specificity, we will build our model around the assumption
 that the precession is induced by the presence of a binary companion, which
 causes the spin axis to precess about the orbital angular momentum vector. 
 However, 
 we emphasize that the families of pulse shape would be the same for any type of 
 spin precession process,  and it would be straightforward to recast this model 
 for  the case of an isolated pulsar's precession. 

\subsection{The Rotating Coordinate System}

The coordinate system we use to model the geometry of the binary pulsar system 
is shown in Fig. \r{basic-geom}.  We keep the same notation as  WT02
where convenient, but choose our coordinates so that the spin vector $\vec K$
and orbital angular momentum vector $\vec J$  of
the pulsar are {\em{stationary}}.   While this coordinate system, which rotates at
the precession rate,  may seem less natural 
than one in which the
bulk of the Universe is stationary (up to cosmological expansion), it will prove 
useful for visualizing the effect of the precession.  A right-handed Cartesian 
coordinate system ($\hat x$, $\hat y$, $\hat z$)
is included in the figure.  The $\hat z$-direction is chosen to be aligned
with $\vec K$, and $\hat x$ is chosen to lie in the plane defined by the
vectors $\vec K$ and $\vec J$.
 %%%%%%%%%%%%%%%%%%%%%%%%%%%%
\begin{figure}
\begin{center}
\includegraphics[height=10cm ]{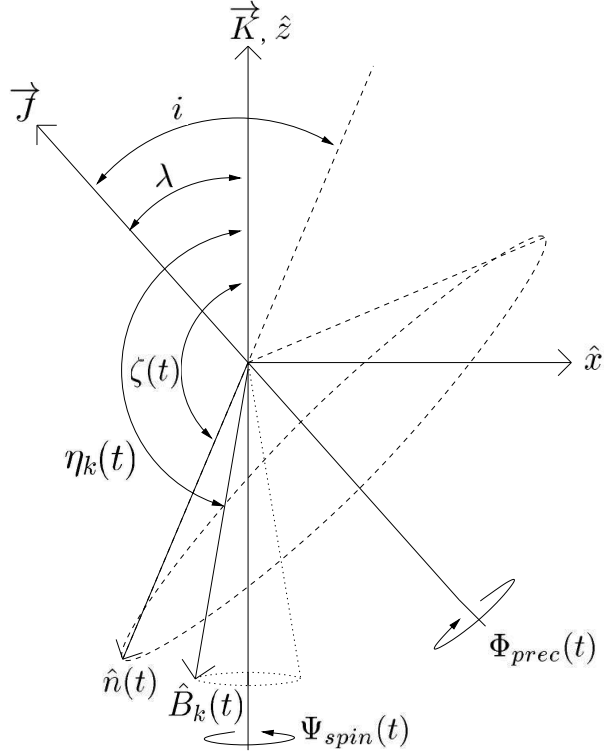}
\caption{Geometry of the binary pulsar in a rotating Cartesian coordinate system 
($\hat x$, $\hat y$, $\hat z$) fixed to the 
pulsar spin vector $\vec K$ and orbital angular momentum vector $\vec J$.  
The $\hat z$-axis of the coordinate system is defined to be parallel to  $\vec K$, while the 
$\hat x$-axis is defined to lie in the plane containing $\vec J$ and $\vec K$ (also the plane of the
page).  Dashes represent the conical trajectory of the pulsar-Earth line of sight $\hat n(t)$ as spin 
precession
carries it around $\vec J$; while dots represent the conical trajectory of a corotating beam-element 
vector   $\hat B_k(t)$
as the pulsar spin carries it around $-\vec K$.   The time-variable vectors $\hat n(t)$ and 
$\hat B_k(t)$   are chosen to be initially in the $\hat x$-$\hat z$ plane.  In accord with standard 
pulsar naming convention,
  $\zeta(t)$ is the (time-variable) spin colatitude of
the pulsar-Earth line of sight.  The colatitude of $\hat B_k(t)$
measured from $\vec K$ is denoted by $\eta_k (t)$, the precession phase is $\Phi_{prec}(t)$ and the spin phase is 
$\Psi_{spin}(t)$.  
See text for additional details.}
\l{basic-geom}
\end{center}
\end{figure}
 %%%%%%%%%%%%%%%%%%%%%%%%%%%%

\subsubsection{Spin Axis Precession}

We consider a model in which the spin vector $\vec
K$, and orbital angular momentum vector $\vec J$ are fixed in space.
The precession of the spin vector $\vec K$ about the orbital
angular momentum vector\footnote{In fact, the spin vector
  precesses about the total angular momentum vector.  However, here
  the spin vector is smaller that the orbital angular momentum vector,
so the total angular momentum vector and the orbital angular momentum
vector are effectively aligned.} $\vec J$ is then equivalent to rotating the rest of
the Universe, including the pulsar-Earth  line of sight $\hat n(t)$, about $\vec J$ (see the
dashed cone upon which $\hat n(t)$ precesses in Fig. \r{basic-geom}).  
The phase of the precession (equivalent to the phase of the cyclical motion of $\hat n(t)$ about $\vec J$)
 is given by $\Phi_{prec}(t)$, where
\be
\l{prec}
\Phi_{prec}(t) = \Omega_{prec}\times (t-t_0).
\ee
Here $t_0$ is a constant chosen so that $\hat n$ lies in the $x$-$z$ plane
at $\Phi_{prec}(t_0)=0$, and $\Omega_{prec}$ is the time averaged spin precession rate of the
pulsar. In Fig. \r{basic-geom}, $i$ is the fixed (orbital)
inclination angle between $\vec J$ and -$\hat n$, $\lambda$ is the fixed spin-orbit
misalignment angle between $\vec K$ and $\vec J$, and $\zeta(t)$ is the 
colatitude  of the line of sight $\hat n(t)$
measured from spin vector $\vec K$, with $\hat n(t)$ describing a cone of half-angle $i$ 
about $\vec J$ on precession timescales.  

The type of spin precession relevant to the 
binary pulsar PSR B1913+16 is relativistic geodetic precession.
\citet{Dam} and \citet{Bar1, Bar2} calculate the rate of such precession for a binary system to be
\be
\Omega_{prec, geodetic} = \frac{1}{2} \left( \frac{G M_{\odot}}{c^3} \right)^{2/3}
\left( \frac{P_b}{2 \pi} \right)^{-5/3} \frac{m_c (4 m_p+3
  m_c)}{(1-e^2) (m_p+m_c)^{4/3}}
\ee
where $m_p$ and $m_c$ are the pulsar and companion masses measured in
units of the solar mass $M_{\odot}$, $P_b$ is the orbital period and
$e$ is the eccentricity.  For the binary system PSR B1913+16 $\Omega_{prec, geodetic}$ is
calculated to be $1.21^{\circ}$yr$^{-1}$  (WT02). 
 [More general expressions for $\Omega_{prec, geodetic}$ in terms of the 
 post-Newtonian
parameterization can be found in \citet{Will}].

\subsubsection{Pulsar Spin}

We define a general vector $\hat B_k(t)$ which corotates with the spinning pulsar at a 
colatitude $\eta_k$ measured from $\vec K$.
For our purposes, $\hat B_k(t)$ can represent the beamed radiation from some particular 
point on
the emission cone.   The effect of the pulsar's rotation is for the vector $\hat  B_k(t)$ to
rotate about $\vec K$.
%While $\hat B_k$ represents {\em{any}} vector corotating with the pulsar, it is useful
%to define a particular vector $\hat \mu$ at colatitude $\alpha$ -- pointing in the direction
%of the magnetic axis of the pulsar.  
%In common with most pulsar emission models, we will assume that the axis of the emission 
%cone will be along $\hat \mu$, with $\alpha$ the colatitude of that axis.  
Fig. \r{basic-geom} shows  a corotating beam vector
%magnetic and emission cone axis 
$\hat B_k(t)$; its  spin about $-\vec K$\footnote{In general,
$\hat B_k(t)$ spins about either $\pm \vec K$, but for PSR B1913+16, it is closer to  $- \vec K$.}
is illustrated with 
dotted lines.  The phase of the rotation $\Psi_{spin}(t)$ is marked on
the figure and is given by $\Psi_{spin}(t)=\omega_{spin} t$ where
$ \omega_{spin}$ is the pulsar spin frequency, which can be written in terms of the pulsar pulse (or spin) period
$P_{spin}$ as $\omega_{spin}=2 \pi/P_{spin}$.

\section{Generating the Observed Pulse Profile}

Having defined the important vectors in \S 2, we can now proceed to generate the
observed pulse profile as a function of time.

\subsection{The Line of Sight and Corotating Beam Element Vectors $\hat n(t)$ and $\hat B_k(t)$}

In considering the precession of the rotating pulsar, we   are interested
in the motion of the pulsar-Earth line of sight $\hat n(t)$, and of a  
vector $\hat B_k(t)$ corotating with the pulsar that represents an emission beam 
element.  These two vectors are the ones that sweep out cones in Fig.\r{basic-geom}. In terms of the
Cartesian coordinates ($x$, $y$, $z$), these vectors can be written
\begin{eqnarray}
\nonumber
\hat n(t)  = \left( \begin{array}{c} x\\y\\z \end{array} \right) &=&
 \left( \begin{array}{c} - \cos \lambda \sin i \cos
\Phi_{prec}(t)+\sin \lambda \cos i\\ \sin i \sin \Phi_{prec}(t)\\-\sin \lambda \sin i \cos
\Phi_{prec}(t)-\cos \lambda \cos i\end{array} \right);\\
\hat B_k(t) &=&  \left( \begin{array}{c} - \sin \eta_k\cos \Psi_{spin}(t)\\- \sin \eta_k\sin
\Psi_{spin}(t) \\\cos \eta_k \end{array} \right).
\end{eqnarray}
The paths swept out by these vectors as they rotate about $\vec J$ and $\vec K$, respectively,
are small circles on the unit sphere centered on the pulsar. Transforming to spherical 
polar coordinates $(\theta,\phi)$  on the unit sphere, where the azimuthal angle 
 in the $(x,y)$-plane is $\theta \equiv \tan^{-1} (y/x)$ 
and the polar angle with respect to the $\hat z$ axis is $\phi \equiv \cos^{-1} z$, we have
\begin{eqnarray}
\nonumber
\hat{n}(t) = \left( \begin{array}{c} \theta\\\phi \end{array}\right) &=&
\left( \begin{array}{c} - \tan^{-1} \left\lbrace \frac{\sin i \sin
    \Phi_{prec}(t)}{(\cos \lambda \sin i \cos \Phi_{prec}(t)-\sin \lambda \cos i}
    \right\rbrace\\ \cos^{-1} \left\lbrace -\cos \lambda \cos i -\sin
  \lambda \sin i \cos \Phi_{prec}(t) \right\rbrace \end{array} \right);\\
\hat B_k(t) &=& \left( \begin{array}{c} \Psi_{spin}(t)\\  \eta_k \end{array}
\right).
\l{sph}
\end{eqnarray}
 In order to
present these trajectories on the page (as we will below), it is necessary 
to project them
from  the above spherical coordinate system onto  a plane with polar
coordinates $(r,\Theta)$, where the radial coordinate\footnote{We choose here $r$ 
to be equal to the supplement of 
$\phi$ rather than $\phi$ itself because the observed
beam is near the $-\vec K$ spin pole for PSR B1913+16. It would be more 
natural to set
$r\equiv \phi$ if the observed beam were nearer the $+\vec K$ spin pole.} is $r \equiv
\pi-\phi$ and the azimuthal angle is $\Theta \equiv \theta$.  They then become
\begin{eqnarray}
\nonumber
\hat n(t) = \left( \begin{array}{c} r\\\Theta \end{array}\right) &=&
\left( \begin{array}{c} \cos^{-1} \left\lbrace \cos i
  \cos \lambda+\cos \Phi_{prec}(t) \sin i\sin \lambda \right\rbrace\\
\cot^{-1} \left\lbrace \cot i \csc \Phi_{prec}(t) \sin \lambda - \cos \lambda
\cot \Phi_{prec}(t) \right\rbrace
\end{array} \right);\\
\hat  B_k(t) &=&  \left( \begin{array}{c} \pi-\eta_k\\  \Psi_{spin}(t) \end{array}
\right).
\l{R2}
\end{eqnarray}
The coordinate system ($\hat r,\hat \Theta$) is chosen such that its
origin corresponds to the point at which  the $-\vec K$  spin axis passes through
the unit sphere.  The small circles swept out by the corotating vector $\hat B_k(t)$ 
on the unit sphere are now circles in the plane, centered on the origin;
while the small circles swept out much more slowly by the line of sight vector 
$\hat n(t)$ are non-circular closed curves (ovals).

\subsection{Generating the Circular Beam from a Set of Beam Elements}

So far we have calculated the geometry of  a single arbitrary point $\hat B_k(t)$, 
which can be taken to be a beam element, corotating  with the pulsar.  
This analysis can be generalized to a conical
beam by intersecting a plane $P_j$
perpendicular to the beam's axis with the unit sphere, giving
a small circle $C_j$ consisting of a {\em{set}} of  $ \hat B_k(t)$ intersecting 
the sphere.  For our ansatz of a circularly symmetric emitting area  the
contours of constant intensity will be concentric circles  on
the unit sphere, corotating with the pulsar.  It is therefore sufficient to consider a
set of concentric circles in order to model our simple circularly
symmetric emitting area.  This picture will be built up by first
considering a single emitting circle, and then many concentric
emitting circles representing the contours of constant intensity.

Since most pulsar emission models center the emission cone on 
the star's magnetic axis  $\hat \mu$ closest to the line of sight, we 
will also define our conal axis to be along $\hat \mu$.  
Because $\hat \mu$ corotates with the pulsar, its trajectory is described by the
equations of a beam vector which we will call $\hat B_0(t)$; i.e.,  $\hat \mu \equiv
\hat B_0(t)$. The spin 
colatitude of the magnetic axis, the angle between $\hat K$ and $\hat \mu$, is 
called $\alpha$.  While our model does not depend on the
beam axis coinciding with the magnetic axis, this specific choice aids in visualizing 
the model and in naming some of the vectors and angles with the conventional 
nomenclature of pulsar astrophysics.

\subsubsection{A Single Circular Emission Cone}

 %%%%%%%%%%%%%%%%%%%%%%%%%%%%
\begin{figure}
\begin{center}
\includegraphics[height=10cm]{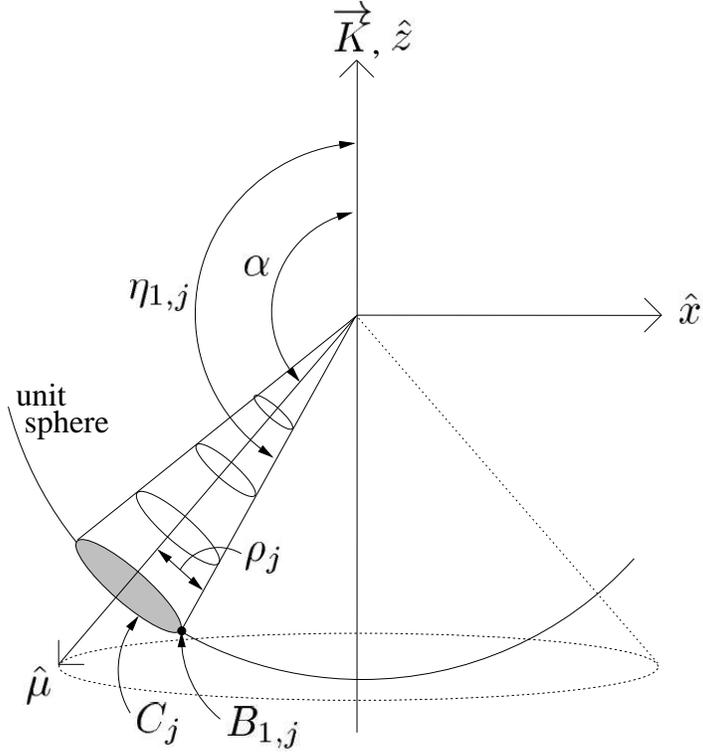}
\caption{An illustration of a single circular emission cone.
 This figure focuses on an entire circular  beam, while Fig. \r{basic-geom} 
showed only a single arbitrary element of the beam, $\hat{B}_k.$  The coordinate system 
is identical to Fig. \r{basic-geom}, with 
$\hat{z}$ fixed to the
  pulsar spin vector $\vec K$ as shown.  A single hollow emission cone, $C_j$,
  centered on the vector $\hat{\mu}$ (which is also the magnetic axis
  in many emission models) is generated by intersecting the unit sphere
  with a (shaded) plane perpendicular to $\hat{\mu}$.  The cone $C_j$ is
  defined by the associated quantities $B_{1,j}$,  $\eta_{1,j}$,
and   $\rho_j$:  $B_{1,j}$ labels the point on $C_j$  that is closest to $-\vec
  K$;  $\eta_{1,j}$ is
  the angle between $\vec K$ and the vector extending from the origin
  to $B_{1,j}$; and $\rho_j$ is the angular radius of $C_j$.     The dotted lines 
  denote the trajectory followed by the beam axis
  $\hat{\mu}$ during a rotation of the pulsar.  Following standard pulsar 
  nomenclature, 
the angle $\alpha$ is the (fixed) spin colatitude of the beam axis.}
\l{beam-figure}
\end{center}
\end{figure}
 %%%%%%%%%%%%%%%%%%%%%%%%%%%%

We will now generate a single infinitesimally thin (i.e., hollow), constant intensity  
emission cone labelled $C_j$ by 
intersecting the unit sphere with a plane $P_{\mu,j}$ defined to be normal to
the cone axis  $\hat \mu$ (see Fig. \r{beam-figure}).  For mathematical simplicity, consider the
moment of time when the beam axis 
is in the $(x,z)$-plane (defined by $\theta=0$) so that
its direction is given by $\hat \mu= \hat x\sin \alpha +
\hat z\cos \alpha$.  Specifying any point in the plane $P_{\mu,j}$, at
this moment, uniquely determines 
it and the small circle $C_j$  (the emission beam) it creates as it
intersects the unit sphere.  We choose this point to be on the surface 
of the sphere in
the  $(x,z)$-plane, where $\theta=0$.  The point is then either the beam 
element closest to or farthest from the spin pole $-\vec K$.  We choose it to be
the closest and label it $B_{1,j}(t) $ with a corresponding
colatitude\footnote{Subscript 1 here denotes the value of this
  quantity at its closest point to $-\vec K$, and does not indicate
  any particular value of $j$.} $\eta_{1,j}$. Note that this small circle of emission, $C_j$, represents a 
constant intensity beam with  angular radius $\rho_j=(\eta_{1,j}-\alpha)$.  
The equation of $C_j$ at this moment is, in our spherical $(\theta,\phi)$ 
system,
\be
\l{circle}
\sin \alpha (\cos \theta \sin \phi-\sin \eta_{1,j})+ \cos \alpha (\cos
\phi -\cos \eta_{1,j})=0;
\ee
which, in terms of the $(r,\Theta)$ polar coordinates,  becomes
\be
\l{circle2}
\sin \alpha (\cos \Theta \sin r-\sin \eta_{1,j})- \cos \alpha (\cos
r +\cos \eta_{1,j})=0.
\ee
An observer at the end of the line of sight vector $\hat n$, with 
colatitude $\phi=\zeta(t)$ from $\vec K$, will see two events in quick succession
as the pulsar's rotation carries the emission beam cone $C_j$ across $\hat n$.
From Eq. \r{circle} we can see that these two events, which we can interpret as the
passage of leading and trailing contours of equal intensity across the line of sight, 
will have the longitudinal (i.e., rotational phase) separation
\be
\l{w}
w_{j}(t)=2 \cos^{-1} \left\lbrace \frac{\cos \rho_j-\cos \alpha
  \cos \zeta(t)}{\sin \alpha \sin \zeta(t)} \right\rbrace
\ee
where it has been assumed that $\hat n$ is effectively static for the
period of one rotation of the pulsar. 

Eq. \r{w} gives $w_j(t)$ in terms of the observer's slowly precessing 
colatitude $\zeta(t)$.  
We can now determine $\zeta(t)$ as a function of the  precession phase
 $\Phi_{prec}(t)$, which is linear in $t$ (see Eq. \r{prec}), by recognizing that
$\zeta(t)$ is just the $\hat \phi$ component of $\hat n(t)$ (see Eq. \r{sph}):
\be
\l{z}
\cos \zeta(t) = -\cos \lambda \cos i- \sin \lambda \sin i \cos \Phi_{prec}(t).
\ee

\subsubsection{Building the Beam from a Set of Circular Emission Cones}

In the previous section we found Eqs. \r{w} and  \r{z} which give the evolution of  $w_j(t)$, the leading-to-trailing 
longitudinal separation of a single $(j^{th})$ circular  intensity contour $C_j$ of the 
beam as a function of $t$.  We will now generalize this notation by
considering $w_j(t)$ to be the time-dependent separations of {\em{multiple}}
 constant intensity contours, each having angular radius $\rho_j$,  where the 
 $j=1,2,3...$ label each contour individually.  By
considering the time evolution of several different $w_j(t)$ simultaneously
for fixed values of $\alpha$, $i$ and $\lambda$, we can build up a
picture of the pulse profile received from the whole emitting cone as
the system precesses.  This will be done in the next section.

\section{The Observable Form of the Pulse Profile}

We have now developed the tools required to analyze and to visualize the 
problem 
of the different types of observed pulse profiles that can be generated as a
binary pulsar undergoes spin axis precession.  Our $(\hat r, \hat \Theta)$ polar
coordinate system centered on the spin axis  is particularly convenient
for this task, since any vector  corotating with the pulsar at colatitude $\eta_k$  
(e.g., a beam element vector $\hat B_k$) will forever traverse a fixed
circular trajectory of radius $(\pi-\eta_k)$ about the $-\hat K$ spin axis at 
angular velocity $\omega=\omega_{spin}$. Meanwhile, the path slowly 
travelled by the precessing line of sight $\hat n$, while also cyclical, will 
be a {\em{deformed}} circle in this coordinate system,  since it is the the 
projection of a circular trajectory onto a lower dimensional reference
plane.  An observer will
see a pulse at those phases of the precession cycle where the line of sight
$\hat n$ intersects some part of the circling beam.  There are a
number of ways that this can occur, leading to a number of qualitatively
different pulse profiles for the observer along $ \hat n$.  These loci
of intersection can be found in the $(\hat r,\hat \Theta)$ coordinate system
using Eqs. (\r{R2}),  the trajectories of  the precessing line of sight 
$\hat  n$ and corotating beam elements $\hat B_k$.

In the following sections, we  present pairs of plots displaying the precessing 
beam from two vantage points.  First, we  display the trajectories 
of the beam and the line of sight for particular geometries in the polar, 
$-\hat K$-centered coordinate system.  However,  {\em{terrestrial}}
observers will not be able to observe such a plot directly; they will
only see a pulse emission profile, changing over the 
precession cycle. Therefore, we also present the form of the pulse profile 
as seen by a terrestrial observer lying along $\hat n$, as a function of 
precession phase. We will call this depiction the ``2DPP''  (two-dimensional 
pulse profile) to distinguish it from conventional 1-dimensional pulse intensity 
profiles. In what follows, we restrict ourselves to geometries
appropriate for the relativistic binary PSR  B1913+16, but the tools developed 
here can be used to visualize variations in the pulse profile of any pulsar 
undergoing spin precession.

\subsection{The Observable Form of a  Single Circular Emission Cone}

To fix ideas we will first consider the observable form of a single corotating emitting cone 
$C_1$, in ($\hat{r}$,  $\hat{\Theta}$) space.  Recall that we showed  $C_j$  in Fig. \r{beam-figure} with its axis
$\hat \mu$ instantaneously ``frozen'' at  some value of $\Theta$.  We also plotted
 the corotating trajectory of the beam axis $\hat B_0=\hat \mu$.  Our
 present purpose is to illustrate how different 2DPPs can be
 generated.  The specific conditions required to separate
the different classes will be found subsequently.  We will restrict
ourselves to  situations producing pulse profiles that approximate those 
observed for the system PSR1913+16 \citep{wtr89,Kramer,Weis, wt05}.
Throughout the 1980's, observations of this system showed that the
pulse profile did not appear to be affected greatly by geodetic
precession.  To generate a relatively unchanging pulse profile during some
period, it is necessary for the line of sight trajectory $\hat n$ to stay at
an approximately  constant radial distance  from the spin axis (the origin of the
($\hat{r}$, $\hat{\Theta}$) plane) during this time,
whilst remaining inside the zone through which the beam circulates.  That is, the oval line of 
sight trajectory must be approximately tangent to that of the emission
beam, in order to minimize precession-induced beam 
shape changes, in concert with the observations made in the 1980's.  In the following
two sections, we describe such line of sight  trajectories as they approach the beam 
circulation zone from the outside and the inside, respectively.

\subsubsection{The Line of Sight Enters into the Beam Circulation Zone
from the Outside}

For the line of sight $\hat n$  to enter the beam circulation zone from the 
outside, we must have  $i>\pi-\eta_{1,2}\sim\pi-\alpha$ (see Fig.   \r{basic-geom}). 
There are two qualitatively different types of such outer trajectories.
Both of these situations are shown together with the resultant
pulse profiles that would be observed  in
Figs. \r{single-cone-outer-entry}a and \r{single-cone-outer-entry}b,
respectively.
The first trajectory in Fig. \r{single-cone-outer-entry}a (the dot-dashed line) is chosen so
that it probes only the outer region of the beam circulation zone.
In this case the line of sight vector $\hat n$ precesses into the
outermost edge of the beam circulation zone,
but never progresses very far into it before precessing out again. 
(Specifically, $\hat n$ never
crosses inside the circle swept out by the beam axis $\hat \mu$.) The second 
trajectory in Fig. \r{single-cone-outer-entry}a (the dashed line) follows much
the same path but critically it precesses further into the  beam circulation
zone before reversing its progress. (In this case, $\hat n$ {\em{does}} cross 
inside of the circle described by $\hat \mu$.) The
effect of this can be clearly seen in the observed pulse profile widths $w$, shown 
as a function of precessional phase in Fig. \r{single-cone-outer-entry}b.
The shallow  outer trajectory, shown as a dot-dashed line, produces a 2DPP
 that varies 
with time in such a way as to produce an oval-shaped contour, whilst the more deeply penetrating dashed
line produces an hourglass shape.  The waist of the hourglass occurs as the line
of sight approaches the {\em{inner}} edge of the emission zone,  the circle
traversed by $\hat B_{1,1}$.  The best fit model
found by \citet{Kramer} lies somewhere between these two models,
where $\hat n$ progresses far enough to create a briefly stationary pulsewidth 
$w$ (as
observed in the 1980's) but not far enough to show the hourglass behavior.
On the other hand, the hourglass-shaped 2DPPs  produced by our {\em{circular}} beam
model are reminiscent of the behavior found more recently by WT02 and WT05,
 leading one to ask if their more complicated beam  model 
is truly necessary. 

 %%%%%%%%%%%%%%%%%%%%%%%%%%%%
\begin{figure}[p]
\begin{center}
\includegraphics[height=9.3cm]{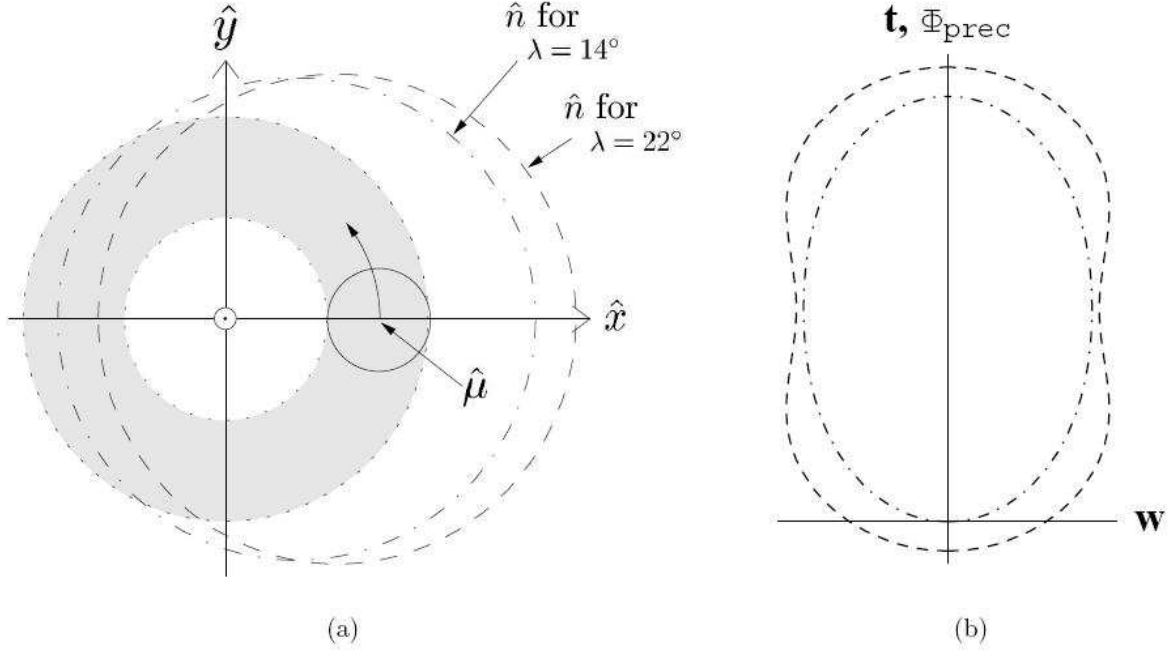}
%\subfigure[]
%%%%%%%{\epsfig{figure=figure3a.eps,height=7.8cm}}
%{     \includegraphics[height=7.8cm ]{f3a.eps}      }
%\qquad
%\subfigure[] 
%%%%%%{\epsfig{figure=figure3b.eps,height=8cm}} 
%{  \includegraphics[height=8cm  ]{f3b.eps}            }
\caption{Line of sight entry points into the beam
circulation zone from the {\it{outside}},
for a {\it{single}} hollow conical beam $C_1$ having radius $\rho_1=10^{\circ}$ and colatitude 
$\alpha=150^{\circ}$; with orbital inclination $i=47\fdg2.$
(a) Spin-axis centered map of beam circulation zone (shaded circular region) 
and precessing lines of 
sight $\hat n$ (ovals) projected onto the $(\hat r, \hat \Theta)$ and $(\hat x, \hat y)$
 plane. The beam $C_1$
(shown as a circle centered on $\hat \mu$) corotates with the pulsar, filling in the shaded zone
in one spin period.
The dot-dashed (dashed) 
oval represents  the trajectory of $\hat n$ for spin-orbit misalignment angle
$\lambda=14^{\circ} \ (22^{\circ})$. (b) The resulting two-dimensional pulse profile
(2DPP), as observed at the end of $\hat{n}$,  showing profile longitudinal 
width, $w$, as a function of precession phase, $\Phi_{prec}$. (Note that $\Phi_{prec}$ is
linear in time:
$\Phi_{prec}=\Omega_{prec} \times (t-t_0)$).  The profile width $w$ 
represents the separation between two pulse components
originating from the leading and trailing portions of the emission cone $C_1$.
Dot-dashed and dashed lines 
are for the same two values of $\lambda$ as in (a).}
\label{single-cone-outer-entry}
\end{center}
\end{figure}
 %%%%%%%%%%%%%%%%%%%%%%%%%%%%

\subsubsection{The Line of Sight Enters into the Beam Circulation Zone
 from the Inside}

Now consider the situation where the line of sight  $\hat n$ enters and 
exits the beam circulation zone through its {\em{inside}} edge (the
one closest to the spin axis; see
Fig. \r{single-cone-inner-entry}) rather than its outside edge as above.  In this case,  
$i< \pi- \eta_{1,1}\sim \pi-\alpha$.  Again, 
there are the oval (hourglass) 2DPP contours 
for trajectories that are allowed to precess slightly (deeply) into the beam
circulation zone.  Indeed, for the case of a single emitting cone, $C_1$,
the inner entry pulse profiles shown in Fig. \r{single-cone-inner-entry}b are quite similar to the
outer entry case shown in Fig. \r{single-cone-outer-entry}b.  We will see below that the outer/inner entry 
degeneracy is lifted when considering multiple emitting cones.

 %%%%%%%%%%%%%%%%%%%%%%%%%%%%
\begin{figure}[p]
\begin{center}
\includegraphics[height=9.1cm]{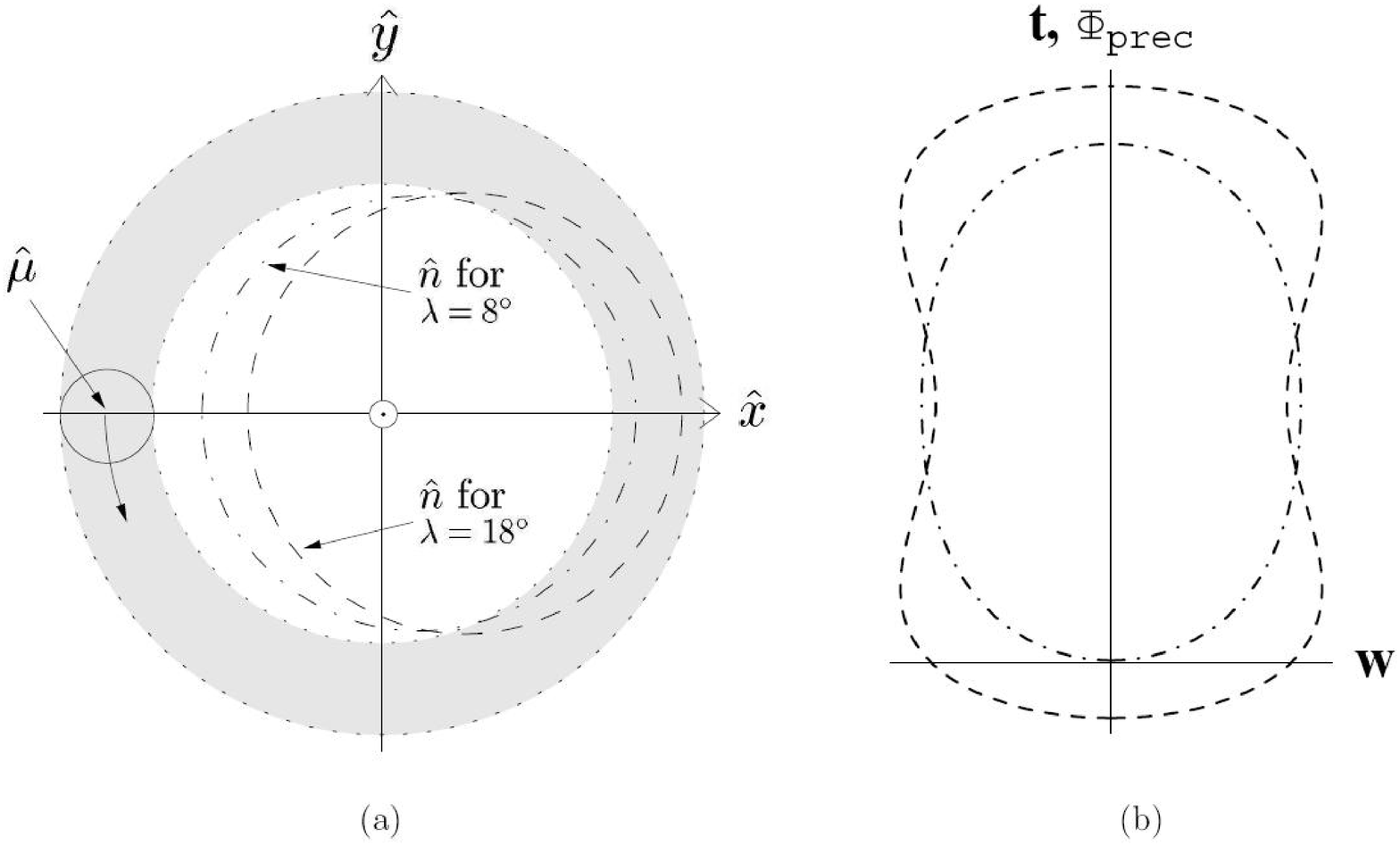}
%\subfigure[]
%%%%%%%%%%%%{\epsfig{figure=figure4a.eps,height=7.7cm}}
%{     \includegraphics[height=7.7cm ]{f4a.eps}         }
%\qquad
%\subfigure[]
%%%%%%%%{\epsfig{figure=figure4b.eps,height=8cm}}
%{      \includegraphics[height=8cm  ]{f4b.eps}              }
\caption{A line of sight entry into the beam
circulation zone from the {\it{inside}},
for a {\it{single}} conal beam $C_1$ having radius $\rho_1=10^{\circ}$ and colatitude  
$\alpha=120^{\circ}$; with  orbital inclination $i=47\fdg2$.   The dot-dashed (dashed) curves are for 
$\lambda=8^{\circ} \ (18^{\circ})$.  See Fig. \r{single-cone-outer-entry} caption
for more details. }
\label{single-cone-inner-entry}
\end{center}
\end{figure}
 %%%%%%%%%%%%%%%%%%%%%%%%%%%%

\subsection{Building Up a Pulse Profile from a Set of  Concentric Emission Cones}

When considering only one infinitesimally thin  emitting cone $C_1$, 
the  differences in 2DPP between the outer and inner line of sight entry cases, illustrated 
respectively in Figs. \r{single-cone-outer-entry}b and
\r{single-cone-inner-entry}b, are difficult to appreciate. However,
when we broaden our considerations to simultaneously include $j$ concentric 
circles $C_j$ (each one representing a distinct 
equal-intensity contour), the observed pulse profiles exhibit a more
complicated and interesting behavior.  We will see that the time evolution of the pulse profile is 
dependent not only on the trajectory of $\hat n$ but also on the radius 
$\rho_j$ of each particular contour $C_j$.  
%(The exact relation will be derived in the next
%section, for the purposes of this chapter we only seek to illustrate
%the qualitative behaviours that can be obtained for the pulse profile).  
%For fixed $i$, $\lambda$ and $\rho_0$ it is then possible for an 
Specifically,  the observer at $\hat n$ would see that some of the $j$  equal-intensity
contours exhibit hourglass-shaped 2DPPs, like the dashed curves of
Figs. \r{single-cone-outer-entry}b and \r{single-cone-inner-entry}b, whilst simultaneously 
seeing others with oval
shaped contours, like the dot-dashed curves in those figures.  
This is particularly interesting as it appears that
observations of PSR1913+16 show exactly this behavior (WT02, WT05).  We
will now show the qualitative difference between the {\em{families}} of 2DPP
contours that are generated for the outer and inner line of sight
entry cases.

Fig. \r{multiple-cone-outer-entry} presents an example of a family of 2DPP contours  generated by an
outer line of sight entry into the beam circulation 
zone.  Fig. \r{multiple-cone-inner-entry} shows an example of an inner entry.
Whilst the form of {\em{single}} contours was shown above to be qualitatively
similar for inner and outer entries of similar depth, 
it is clear from Figs. \r{multiple-cone-outer-entry} and \r{multiple-cone-inner-entry} 
that {\em{families}} of
concentric contours are quite different in the outer and inner entry cases.  
In Fig. \r{multiple-cone-outer-entry}b the
outer 2DPP contours are ovals while the inner ones are
 hourglass-shaped.  Figure \r{multiple-cone-inner-entry}b exhibits exactly the opposite behavior.

The best fit model of \citet{Kramer} represents a shallow outer line of sight entry 
with\footnote{By Kramer's definition of $\alpha$, this condition translates to $i>\alpha$.} 
$i>\pi-\alpha$, as
illustrated by the dot-dashed curve in Fig. \r{single-cone-outer-entry}.
 Whilst Kramer's best fit model does accurately reproduce the
observations for a {\it{single}} contour of PSR B1913+16 (the contour
of peak intensity, as measured by the Arecibo and Effelsberg
telescopes up through  the mid-1990s), it does not appear to match the
observations  when a whole family of concentric
contours are considered\footnote{This is due to Kramer's
model belonging to our `outer entry' class of models.}  (WT02, WT05).  
However, the simple circularly symmetric, {\em{multiple}} 
cone model  considered here appears able to produce the same approximate shape as the 
latter observations  (i.e,
oval inner contours and hourglass-shaped outer ones (see Fig. 
 \r{multiple-cone-outer-entry})).
 %%%%%%%%%%%%%%%%%%%%%%%%%%%%
\begin{figure}[p]
\begin{center}
\includegraphics[height=7.5cm]{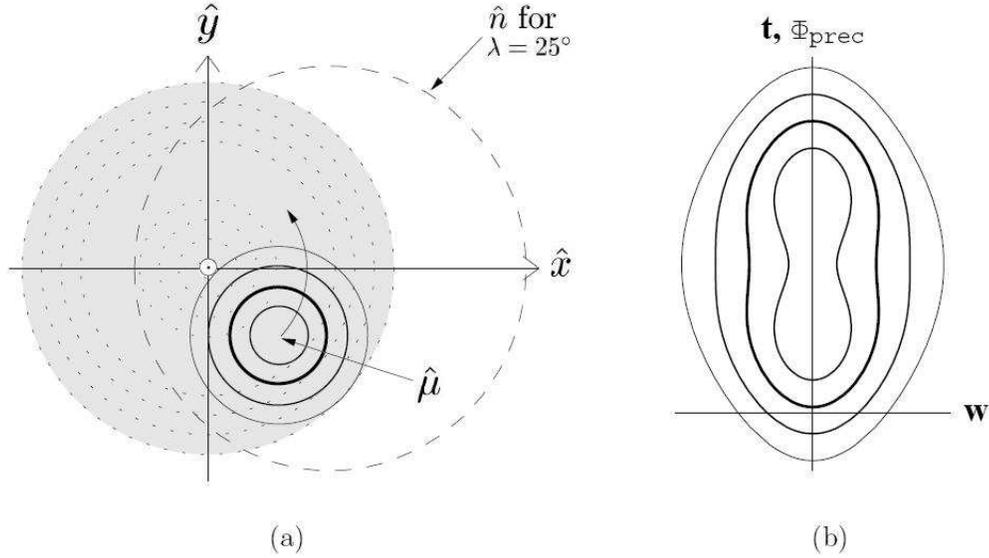}
%\subfigure[]
%%%%%%{\epsfig{figure=figure5a.eps,height=6cm}}
%{      \includegraphics[height=6cm]{f5a.eps}         }
%\qquad
%\subfigure[]
%%%%%{\epsfig{figure=figure5b.eps,height=6cm}}
%{       \includegraphics[height=6cm ]{f5b.eps}       }
\caption{A line of sight entry into the beam circulation zone from the
 {\it{outside}}, for a {\it{multiple}} cone model with conal radii  $\rho=6^{\circ}, 10^{\circ}, 14^{\circ}$
  and $18^{\circ}$; and $\lambda=25^{\circ}$,
  $\alpha=160^{\circ}$ and $i=40^{\circ}$.
 (a) Spin-axis centered map of beam circulation zone 
  (shaded circular region);  and  the trajectory of the  precessing line of sight 
  vector $\hat{n}$ (dashed line)   projected onto the $(\hat r, \hat \Theta)$ and  
  $(\hat x, \hat y)$ plane. The
  concentric solid circles centered on $\hat \mu$ depict the instantaneous position 
  of the hollow, circular emission beams, with the line thickness representing the intensity 
  of each one.  The beams corotate with the pulsar, filling in the shaded ``beam circulation zone"
   in one spin period.
  The dotted lines represent the spin trajectories of inner and outer edges of the various conical
  beams.   (b) The resulting two-dimensional pulse profile
(2DPP), as observed at the end of $\hat{n}$, showing  profile longitudinal 
width  $w$ as a function of precession phase, $\Phi_{prec}$. 
(Note that $\Phi_{prec}$ is linear in time:
$\Phi_{prec}=\Omega_{prec} \times (t-t_0)$).  Each closed curve corresponds to a separate
  circular emission cone.  The profile width $w$ across a particular contour
represents the separation between two pulse components
originating from the leading and trailing portions of each emission cone.    Note that inner contours
are hourglass-shaped, while outer ones are oval.
See Fig. \r{single-cone-outer-entry} caption for more details. }
\label{multiple-cone-outer-entry}
\end{center}
\end{figure}
 %%%%%%%%%%%%%%%%%%%%%%%%%%%%
  %%%%%%%%%%%%%%%%%%%%%%%%%%%%
\begin{figure}[p]
\begin{center}
\includegraphics[height=9.1cm]{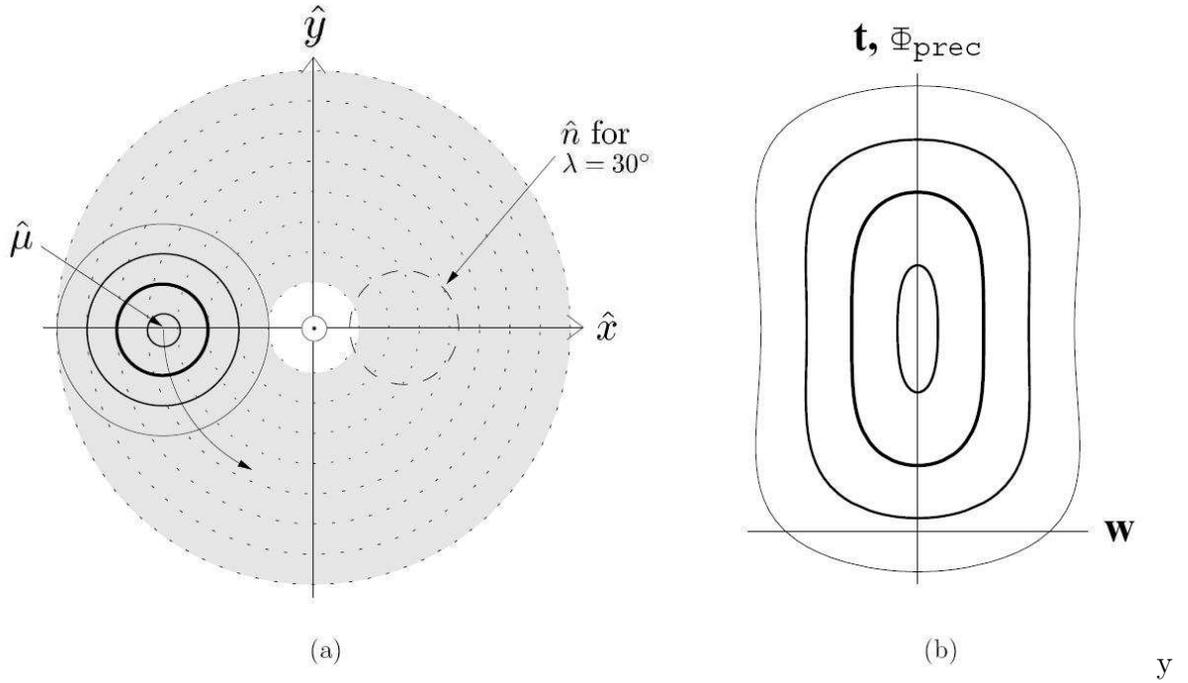}
%\subfigure[]
  %%%%%%%%%{\epsfig{figure=figure6a.eps,height=8cm}}
%{     \includegraphics[height=8cm ]{f6a.eps}        }
%\qquad
%\subfigure[]
  %%%%%%%%{\epsfig{figure=figure6b.eps,height=8cm}} 
%{    \includegraphics[height=8cm ]{f6b.eps}        }
y\caption{A line of sight entry into the beam circulation zone from the
 {\it{inside}}, for a {\it{multiple}} cone model with conal radii $\rho=5^{\circ}, 15^{\circ}, 25^{\circ}$
  and $35^{\circ}$; and   $\lambda=30^{\circ}$,
  $\alpha=130^{\circ}$ and $i=18^{\circ}$.  Note that the 2DPP shown in (b) resemble
  those of PSR B1913+16 (WT02, WT05), with inner ovals and outer hourglasses.  See
  the captions of Figs. \r{multiple-cone-outer-entry} and
  \r{single-cone-outer-entry} for more details.}
\label{multiple-cone-inner-entry}
\end{center}
\end{figure}
 %%%%%%%%%%%%%%%%%%%%%%%%%%%%

\subsection{Conditions for Oval- and Hourglass-shaped Two-dimensional Pulse Profiles}

It was seen in the previous section that it is possible to generate
both oval- and hourglass-shaped 2DPPs from a circularly
symmetric emitting region.  We will now derive the specific
conditions necessary to produce both oval and hourglass shapes
together in the same two-dimensional pulse profile.

To find the precession
phases $\Phi_{prec}$ at which the pulsewidth  $w_j$ of beam $j$ 
is unchanging we must find the point at which
\be
\l{deriv}
\frac{d w_j}{d \Phi_{prec}}=0
\ee
for a smooth function $w_j(\Phi_{prec}).$\footnote{In terms of Figs. 
[\r{single-cone-outer-entry}b-\r{multiple-cone-inner-entry}b], we are 
finding the phases at which the  $j^{th}$ 2DPP contour is
vertical.} This condition occurs when
\be
\qquad
\sin \Phi_{prec} (\cot \alpha +\cos \rho_j \csc \alpha (\cos i \cos
\lambda+\cos \Phi_{prec} \sin i \sin \lambda))=0.
\ee
For an oval-shaped 2DPP, this expression will vanish only when $\sin
\Phi_{prec} =0$; whereas an hourglass profile will have this quantity vanish
two extra times when
\be
\cos \Phi_{prec} = -\cot i \cot \lambda- \cos \alpha \csc i \csc \lambda \sec
\rho_j.
\ee
The condition that the profile be hourglass-shaped is then the
condition that the right hand side of the above equation lie between
$-1$ and $1$.  This is satisfied if $\rho_j$ lies in the range
\be
\l{condition}
\cos (i+\lambda) < -\cos \alpha \sec \rho_j < \cos (i-\lambda)
\ee
if $\sin i>0$, as it is for PSR B1913+16.  For a specified intensity
contour and binary geometry it is now straightforward to see if a
2DPP contour will be an oval or an hourglass.

Furthermore, Equation (\r{condition}) can be used to establish
whether a family of contours will produce hourglass shapes for its
smallest or largest radii - that is, whether its pulse profile looks like
Fig. \r{multiple-cone-outer-entry} or Fig. \r{multiple-cone-inner-entry}.  
To see this we first note that $\rho$ must lie
within the range $0$ to $\pi/2$, in order that the emitting region should
cover less than the entire surface of the pulsar.  The quantity $\cos
\alpha \sec \rho$ then increases in magnitude as $\rho$ becomes
larger (for fixed $\alpha$)\footnote{We restrict ourselves here to
considering $\alpha>\pi/2$, as appropriate for PSR B1913+16.}.

\subsubsection{Small Hourglasses}

We first consider the conditions necessary to produce hourglass-shaped 
{\em{inner}} 2DPP contours; i.e, those that are hourglasses as 
$\rho \rightarrow 0$, so that $\sec \rho
\rightarrow 1$. Taking this limit of Eq. \r{condition} we find that if $\alpha$ lies in the range
\be
\l{condition2}
\cos (i+\lambda) < -\cos \alpha < \cos (i-\lambda),
\ee
then the innermost profile contours  will be
hourglass-shaped.  Now, if $(-\cos \alpha)$ satisfies
Eq. \r{condition2}, and is
also sufficiently close to $\cos
(i-\lambda)$, then  the contours with larger $\rho$ will be ovals even while
the inner ones remain hourglass-shaped.  

\subsubsection{Small Ovals}

For the smallest 2DPP contours to be oval-shaped, $\alpha$
must satisfy either
\be
-\cos \alpha < \cos (i+\lambda) \qquad
\ee
 or
 \be
 \qquad -\cos \alpha > \cos (i-\lambda).
\ee
The first of these small-radius oval-contour conditions allows for the
possibility of larger radii simultaneously producing hourglass-shaped contours, and is a necessary
condition to generate a profile of the form shown in 
Fig. \r{multiple-cone-inner-entry}.  The second
of the above bounds signifies situations in which all of the contours
produce ovals.

\section{Confrontation with the PSR B1913+16 Data}

Having presented a way of visualizing the effects of
precession, and having investigated some of the pulse profiles that can be
achieved from a simple circularly symmetric emitting region, we will
now confront our ideas with observational data.  This
will allow us to consider the degree to which deviations from
this simplest model are required in order to explain the observations of
PSR B1913+16.  By finding the best fitting parameters for this model,
we will also have found plausible starting points about which more
complicated, future studies can focus.

\subsection{Data Acquisition and Preliminary Analyses} 

All data for this study were collected at $\lambda\sim21$ cm at Arecibo Observatory. \citet{taylor89} present 
descriptions of the pulsar observing systems employed. The details of the process leading to a final  
 ``session-average" profile for each of twenty-three two-week observing sessions from 1981 to 2003 are
 given in WT02. Those authors and WT05 analyzed the same data set studied here.  
 
The data exhibit the 
 following general features. The double-peaked pulse profile exhibits a $\sim 1 \% /$yr decline
 in the ratio of leading to trailing peak intensity, as first discovered by \citet{wtr89}. Until the mid-1990s,
 no other changes were detected, indicating that the $\sim$middle of a hollow conical beam was 
precessing across the observer's line of sight.  The intensity ratio
 change was ascribed to locally ``patchy'' structure in the conical
 beam precessing across the line of sight.  \citet{Kramer} was the
first to discover a narrowing of the separation between the
 two principal pulse component peaks, indicating that the 
 center of the beam was finally precessing away from  the line of sight.  
 
 All subsequent observations indicate that the profile narrowing continued, but with
 some interesting twists.  The 2DPPs of WT02 and WT05 show that whilst the profile peaks
 moved together and the saddle region between them filled in, the outermost intensity contours 
 did not converge over time and may have even diverged.  It is our
 intention to use the new understanding delineated in \S4 to 
search for possible circular beam solutions that could account for
 these observations.

\subsection{Model Fits}

The preliminary stages of this investigation follow the procedures of WT02 and WT05.
Each session-average profile was split into even and odd parts, and subsequent analyses
focused only on  the
even parts, under the assumption that the odd parts represented local, ``patchy'' structure not
relevant to overall beam modelling. The pulsewidth $w$ was then determined for each of fourteen
intensity levels in all of the 23 session-average even profiles.

We then diverge from the procedures of WT02 and WT05 by fitting
these data to our circularly symmetric model, rather than to their elongated beam
model.  We fix the  sine of the orbital inclination,  $\sin i$,  at 0.734 [from the timing measurements 
of \citet{taylor89}], and fit for four parameters:  The colatitude of the magnetic
axis $\alpha$, the spin-orbit misalignment angle $\lambda$ (see
Fig. \r{basic-geom}), the 
precession epoch $T_0$ and the overall scale factor $s$. [See WT02 for further discussion
of these parameters.]   \citet{Kramer} fitted for quantities similar to these 
parameters with his single contour model, and data on the separation between 
the profile peaks only.  While the fitting process alone cannot distinguish between four
degenerate solutions, the earlier work of these authors shows that one of the four is
favored for other reasons.  In what follows, we will focus only on this one, which in all cases
has $i=47\fdg20$.

% \citet{Weis} also fitted for these
%parameters, and two additional ones describing their generalized beam
%shape, with the data from a number of contours of constant intensity.  We will
%now attempt to fit the same multiple emission contours here, without
%resorting to a modified beam shape. 

\subsubsection{Fitting to the Late Precession Phase}

Let us first consider focusing the fit on  the part of the two-dimensional pulse
profile immediately before the
pulsar beam precesses entirely out of view (i.e., at late precession phases).  
In this case, it can be seen
immediately  that geometries in which
the precessing line of sight vector, $\hat n$, enters and exits the emission beam on the {\it{outside}}
(as in Fig. \r{multiple-cone-outer-entry}b ) are not good candidates to explain the 
observations.  Conversely, the late-precession-phase  2DPP
generated from the line of sight entering and exiting the emission beam from the {\it{inside}},
shown in Fig. \r{multiple-cone-inner-entry}b, match the  observations much better.  With these
2DPPs we have the inner contours converging while the outer
%%%%%%%%%%%%%%%%%%%%%%%%%%%%
\begin{figure}
\begin{center}
\includegraphics[height=15cm,angle=180]{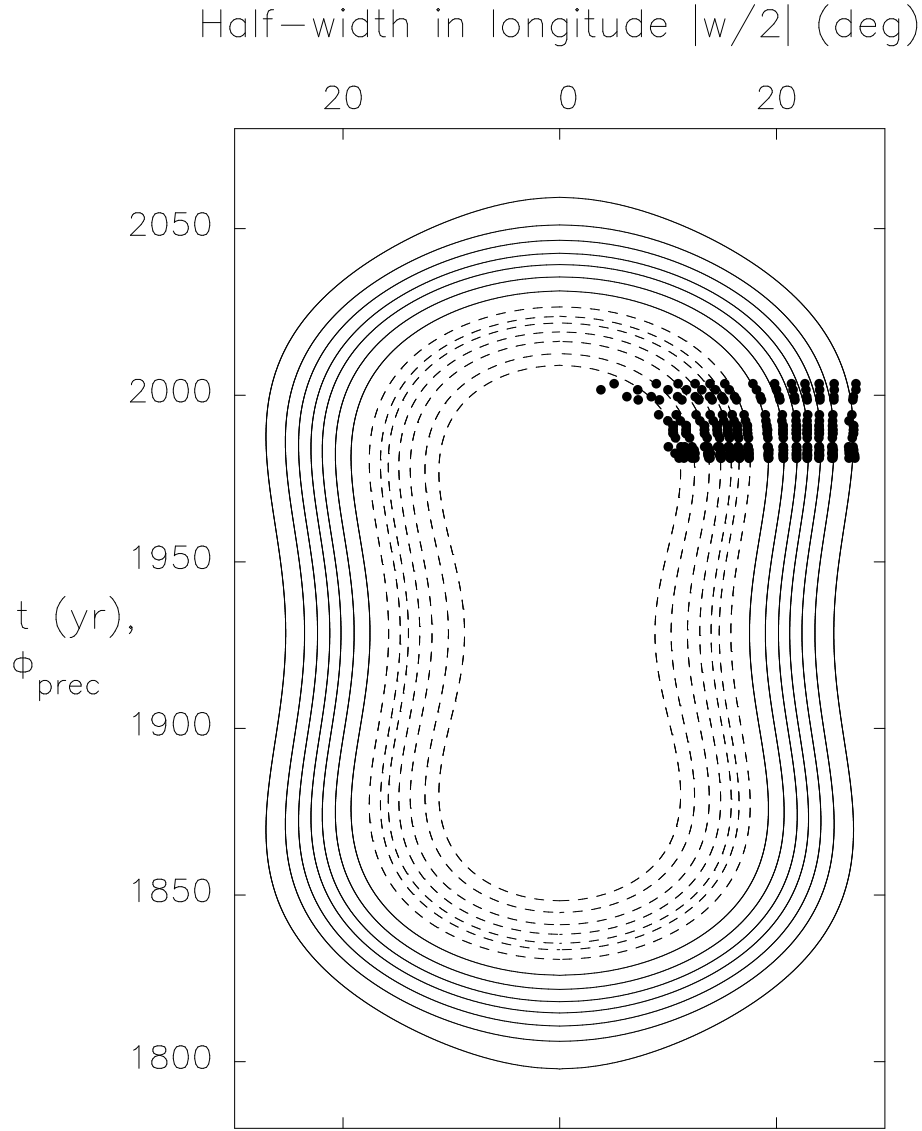}
\caption{Late precession-phase fit.  Fitting of the two-dimensional pulsewidth data of WT02 and WT05 
to late precession phases of a  a model with 
$\alpha=123\fdg7$ and $\lambda=14\fdg5$.  Each equal-intensity contour results from 
emission from a circularly symmetric
conical beam.   The intensity rises from one  inner dashed contour to the next outer one; and then 
declines as one moves outwards among the solid contours.  The vertical axis is calibrated in years,
but can also be considered to be precessional phase since $\Phi_{prec}$ is linear in time.  
See text for additional details.}
\l{end-result}
\end{center}
\end{figure}
\begin{figure}
\begin{center}
\includegraphics[height=15cm,angle=180]{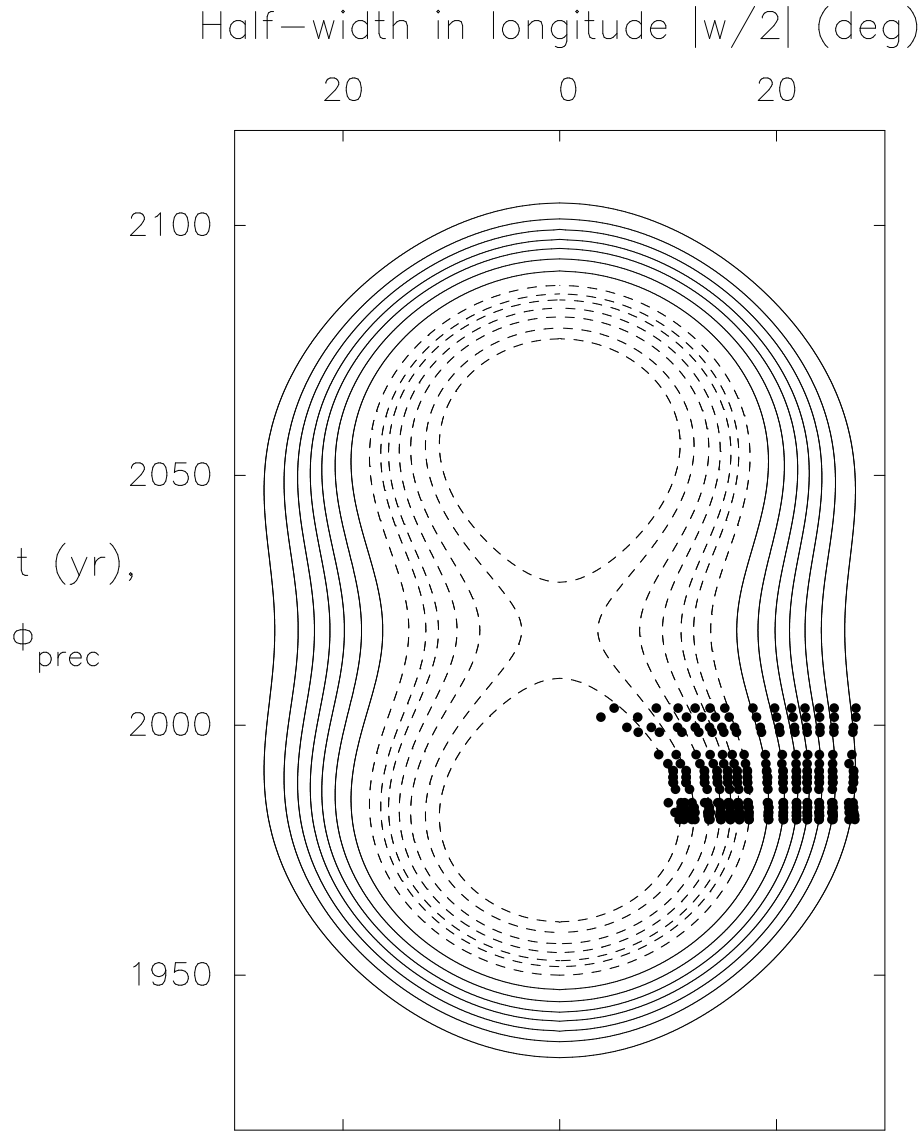}
\caption{``Pre-waist'' fit to 2DPPs. In this case, 
$\alpha=142\fdg8$ and $\lambda=17\fdg2$.  Each equal-intensity contour results from 
emission from a circularly symmetric
conical beam.  
See Fig. \ref{end-result} and  text for additional details.}
\l{waist-result}
\end{center}
\end{figure}
 %%%%%%%%%%%%%%%%%%%%%%%%%%%%
ones remain approximately stationary, in keeping with the observations of PSR1913+16. 
 There is also the possibility of achieving a
period shortly before this in which {\it{all}} the contours remain $\sim$stationary, as was observed by
\citet{wtr89}.  Such a period
of stationarity is achieved by making the outer contours only mildly
hourglass shaped, so that they appear as almost ``pill'' shaped.  We display our attempts to find a
fit in this ``late precession phase'' scenario in Fig. \r{end-result}.  We find an
approximate match when  $\alpha=123\fdg7$ and $\lambda=14\fdg5$.  However,
%, and $s=1.966$. 
%1780+297=2077
as indicated by the systematic deviations of data from the model illustrated in
the figure, and quantified by a large $\chi^2$, the fit is not
particularly good.  In fact, this particular configuration does not
appear as a stable solution of the fitting program.
 
 \subsubsection{Fitting to the Pre-waist Precession Phase}

A second place at which to attempt a fit is near the `waist' of the model 2DPP 
hourglass (i.e., near the middle phases of the precession cycle).  In this case it is 
the profiles generated just  {\it{before}} the midpoint 
of  an  {\it{outer}} line of sight  precession cycle 
(see Fig. \r{multiple-cone-outer-entry}b) that best match the 
observed ``inner-in'' and ``outer-out'' contours; whilst the {\it{inner}} entry/exit 
2DPP (Fig. \r{multiple-cone-inner-entry}b) display the incorrect behavior 
near these precession phases.  
%These
%profiles have hourglass shapes for their inner contours, and ovals for
%their outer ones.  Therefore, at the `waist' of the 2DPP we have a
%period in which the inner contours are converging, while the outer
%ones diverge. 
% In this situation it is easier
%to obtain outer contours that are coming together slowly, with respect
%to the inner ones, but in order to make the contours come
%together at the appropriate rate it is necessary to make them
%strongly hourglass-shaped.  This makes it more difficult to achieve the
%required period of stationarity. 
Unfortunately,  the recently observed convergence of inner contours requires
such strong hourglass shapes that it is difficult to produce the
required period of stationarity with this class of fit.  We display our attempt to find a
match in this  ``pre-waist'' scenario in Fig. \r{waist-result}.  We find an
approximate fit with $\alpha=142\fdg8$ and $\lambda=17\fdg2$.
%, and $s=1.420$. 
Again, however, the
%%%%%approximate fit with $\alpha=143\fdg6$,
%1720+297=2017
%$\lambda=17\fdg6, t_0=2017$, and $s=1.395$. Again, however, the
fit is poor.

 \subsection{Discussion}
Clearly these two results are not convincing enough to offer a complete explanation of
the observed pulse profile from PSR1913+16 by themselves; deviations
from this minimal model, such as those considered by WT02, should
be considered in order to find a better fit.  However, the results of
this study do show us where in parameter space we
can start future searches with more complicated beam shapes.  By
searching in the vicinity of our approximate fits, we expect that the
observed pulse profiles may be explained with only minimal
deviations from this maximally symmetric model.  
%The requirement of
%deviations from a perfectly circular beam should not come as a
%surprise, as in order to obtain the symmetric component of the
%observations that we have been fitting to, it has been necessary to
%disregard an anti-symmetric component.  Such a component cannot be
%explained within the context of this model without the additional assumption of
%a locally patchy structure.

A number of other pulsars exhibit long-term pulse profile changes that may be 
precession-induced, and which could thus benefit from the analysis
techniques developed here.  The candidates can be divided into two
classes: First, binary pulsars 
like B1913+16 that are expected to undergo geodetic precession due to relativistic
spin-orbit coupling; and second, isolated pulsars undergoing free precession caused by
a body asymmetry.  

Among the binary pulsars, B1913+16 currently has the most detailed and longest duration 
pulse shape measurements. Precession-induced pulse shape changes have also been detected
in B1534+12 \citep{sET04}, J1141-6545 \citep{hET05}, J1906+0746
\citep{lET06,kET07}, and J0737-3039B \citep{bET05}.  However,
with the exception of the last pulsar, the trends have
principally been roughly linear in time\footnote{Recent progress indicates
  that this may not be the case with J1141-6545 \citep{Krampriv}.}. 
The final listed object 
is a member of a complicated double pulsar system, where magnetospheric interactions between 
the two stars are operating along with orbital and spin axis precession.  Hence our model will yield no 
unique insights into these systems, at least until data are accumulated across a significant portion of the 
precession cycles.

Among the isolated pulsars showing pulse shape changes, most sources again seem to reveal
 long-term quasi-linear trends rather than periodic behavior.  [See \citet{wET07} for a review.]  The most
 promising object is B1828-11 \citep{sET00,sET03}, which shows periodic pulse shape changes on
 timescales of $10^{2-3}$ days.  However, its signature exhibits a
 double-narrowing, a feature that is difficult to
 reproduce with a model of the type considered here.  This suggests
 that B1828-11 manifests significant deviations from the highly
 symmetric configurations we have been considering.

\section{Conclusions}

We have investigated the possible pulse profiles that can be generated
from a precessing pulsar which emits a simple circularly symmetric beam.  
We have found that a coordinate system fixed to the spin and orbital angular momentum 
vectors provides a useful means of visualizing the precession process and its
observable consequences.  We showed that a variety of different two-dimensional
pulse  profiles (2DPP) can be
generated from one or more {\it{circular}} beams, including profiles where the contours of 
constant intensity can
appear as either ovals or hourglass shapes.  Furthermore, we showed that it
is possible to create 2DPP which are a combination of
ovals and hourglasses, with the hourglass behaviour occurring at either
large or small pulse widths depending on the geometry of the system.

This work has direct application to the determination of
the geometry of binary systems undergoing spin precession, such as PSR 1913+16.  
Best fit models have
previously been constructed for this system.  \citet{Kramer} found a best
fit circular model for a single intensity contour.  This model, whilst accurately modelling
the peak of the profile, does not appear to produce a family of
contours which fits the observational data of WT02 and WT05, while
simultaneously maintaining the
assumption of a circularly symmetric emitting region.  Subsequent
studies by WT02 and WT05 show that by giving up circular symmetry 
(specifically by positing an hourglass-shaped beam),
it is possible to reproduce a family
of contours that fits the observations well.

The results found here show that simple circularly
symmetric emitting regions can generate pulse profiles which
have the same qualitative form as the observational data of WT02 and WT05
- oval shaped inner contours and outer contours which come together at
a later time, and are possibly even hourglass shaped.  A more detailed
investigation, however, has shown that this simple model is not
sufficient to fit the observations completely.  Never the less, the
approximate fits we have found suggest that only minimal deviations
from a perfectly circular beam  may be sufficient to explain
the data.

The requirement of deviations from circular symmetry should not come as
a complete surprise, as we know the full pulse profile contains an
asymmetric component, that cannot be easily accounted for with a
circular beam.  We expect the inclusion of the anti-symmetric
component in future analyses should give important information about
the detailed shape of the beam.

We have shown that it is not necessary to deform the emission beam to be the same shape
as the observed pulse profile, in order to produce (at least qualitatively) the required
shapes.  Furthermore, this investigation has shown the
most likely places to start looking for fits with more complicated
beam shapes, if only minimal deviations from the circular model are
desired.

We expect that there will be a great many ways in which to deform
the beam shape so as to produce better fits to the data.  For example, one
may consider circular contours of constant intensity that are not
centered on exactly the same point; or, one may wish to deform each of
the circles by squashing and stretching them in various ways.  We
expect that it will be a matter of some difficulty to show which
deviations are really the most probable, but expect that it will be
useful to be guided by Ockham's razor, and consider the most minimal
deviations to be the most likely.

Of course, future observations will be of great use in determining the
true parameters of PSR 1913+16:  Different models predict that the
pulsar will precess in and out of our line of sight in different periods of time.
(Compare for example, the future behavior of the pulsar as illustrated
in Fig. \ref{end-result} {\it{versus}} Fig. \ref{waist-result}.)
As we continue to build up  a more complete picture of the two-dimensional
pulse profile, it will be a considerably easier task to find the
true geometry of the system.

We have also surveyed the observations of possibly precession-induced pulse shape changes
in other pulsars.  Most of these observations are not yet sufficiently detailed to benefit from our 
analysis technique.  As more data are gathered in the future, these  procedures will prove 
useful.

\vspace{40pt}

\leftline{\bf Acknowledgements}

We would like to thank J. Barrow and M. Kramer for helpful discussions,
and S. Vigeland for assistance with programming and visualization.
TC acknowledges the support of the Lindemann trust.  JMW has been supported by
NSF Grant AST 0406832.  Arecibo Observatory is operated by Cornell University under 
cooperative agreement with the NSF.


\begin{thebibliography}{}

\bibitem[Barker \& O'Connell(1975a)]{Bar1} Barker, B.~M., \& 
O'Connell, R.~F.\ 1975, \prd, 12, 329 

\bibitem[Barker \& O'Connell(1975b)]{Bar2} Barker, B.~M., \& 
O'Connell, R.~F.\ 1975, \apjl, 199, L25 

\bibitem[B\"orner, Ehlers \& Rudolph (1975)]{Borner} B\"orner, G.,
  Ehlers, J., \& Rudolph, E.\ 1975, A\&A, 44, 417

\bibitem[Burgay et al.(2005)]{bET05} Burgay, M., et al.\ 
2005, \apjl, 624, L113 

\bibitem[Damour \& Ruffini(1974)]{Dam} Damour, T., \& Ruffini, R. 1974, 
	Academie des Sciences Paris Comptes Rendus Serie Sciences 
	Mathematiques, 279, 971

%\bibitem[Esposito \& Harrison(1975)]{Esp} Esposito, L. W.  \& Harrison,  E. R., 1975, \apj,196, L1 

\bibitem[Hari Dass \& Radhakrishnan(1975)]{hari75} Hari Dass, 
N.~D., \& Radhakrishnan, V.\ 1975, \aplett, 16, 135 

\bibitem[Hotan et al.(2005)]{hET05} Hotan, A.~W., Bailes, M., 
	\& Ord, S.~M.\ 2005, \apj, 624, 906 

\bibitem[Kasian et al.(2007)]{kET07} Kasian, 
L.~E., \& PALFA consortium. \ 2007, ArXiv e-prints, 711,  arXiv:0711.2690 

\bibitem[Kramer(1998)]{Kramer} Kramer, M., 1998, \apj, 509, 856

\bibitem[Kramer, private communication (2007)]{Krampriv} Kramer, M., private
  communication 2007

\bibitem[Lorimer et al.(2006)]{lET06} Lorimer, D.~R., et al.\ 
2006, \apj, 640, 428 

\bibitem[Stairs et al.(2000)]{sET00} Stairs, I.~H., Lyne, A.~G.,  \& Shemar, S. L.\ 2000, 	Nature, 406, 484 

\bibitem[Stairs et al.(2003)]{sET03} Stairs, I.~H., 
	Athanasiadis, D., Kramer, M., \& Lyne, A.~G.\ 2003, Radio Pulsars, ASP Conf. Ser. 302, 
	Ed. M. Bailes, D. J. Nice, \& S. E. Thorsett. San Francisco: ASP, 249 

\bibitem[Stairs et al.(2004)]{sET04} Stairs, I.~H., Thorsett, 
	S.~E., \& Arzoumanian, Z.\ 2004, Physical Review Letters, 93, 141101 

\bibitem[Taylor \& Weisberg(1989)]{taylor89}  Taylor, J. H. \& Weisberg, J. M. 1989, ApJ, 345, 434

\bibitem[Weisberg, Taylor, \& Romani(1989)]{wtr89} Weisberg, J.~M., 
Romani, R.~W., \& Taylor, J.~H.\ 1989, \apj, 347, 1030 

\bibitem[Weisberg \& Taylor(2002)]{Weis} Weisberg, J. M.  \& Taylor, J.H. 2002, \apj,
	576, 942  (WT02)
	
\bibitem[Weisberg \& Taylor(2005)]{wt05} Weisberg, J.~M., \& 
Taylor, J.~H.\ 2005, Binary Radio Pulsars, ASP Conf. Ser. 328, 
Ed. F. A. Rasio \& I. H. Stairs. San Francisco: ASP,  25 (WT05). E-print available at
http://arxiv.org/abs/astro-ph/0407149 

\bibitem[Weisberg et al(2007)]{wET07} Weisberg, J.~M., et al,\ 2007, astro-ph/07xxxxx

\bibitem[Will(1993)]{Will} Will, C. M. 1993, Theory and Experiment in Gravitational
  Physics, Cambridge:  Cambridge U. Press.


\end{thebibliography}
\end{document}